\documentclass[12pt]{article}
\pdfoutput=1

\usepackage{amsmath,amssymb,amscd}
\usepackage{listings}
\usepackage{caption}
\usepackage{dsfont}
\usepackage{slashed}
\usepackage{color}
\usepackage{ulem}
\usepackage[pdftex]{graphicx}
\usepackage{epstopdf}
\usepackage{subfigure}
\usepackage{epsfig}
\usepackage{listings}
\usepackage{caption}
\usepackage{cite}
\usepackage{multirow}
\usepackage{booktabs}
\usepackage{comment}
\usepackage{lineno}
\usepackage[
    colorlinks=true,  
    linkcolor=blue,   
    filecolor=magenta, 
    urlcolor=blue,    
    pdftitle={你的文档标题},  
    pdfauthor={你的名字}      
]{hyperref}
\usepackage{cleveref}  
\usepackage{placeins}  

\newcommand{\beq}{\begin{equation}}
\newcommand{\eeq}{\end{equation}}
\newcommand{\nbea}{\begin{align*}}
\newcommand{\neea}{\end{align*}}
\newcommand{\nbeq}{\begin{equation*}}
\newcommand{\neeq}{\end{equation*}}

\usepackage{multirow}
\usepackage{array}
\newcolumntype{M}[1]{>{\centering\arraybackslash}m{#1}}
\newcolumntype{N}{@{}m{0pt}@{}}

\newcommand{\BHABHA}{\rm e^+ e^-\to e^+ e^- }
\newcommand{\BHABHAG}{\rm e^+ e^-\to e^+ e^-(\gamma) }

\newcommand{\dd}{\mathrm{d}}

\begin{document}


\pagestyle{plain}

\baselineskip=21pt

\begin{center}

{\large {\bf New test on contact interactions in the data at $\sqrt{s}$ 130-207GeV by Bhabha scattering process $\BHABHA$}}

\vskip 0.3in

\bf Zhikun Xi \textsuperscript{a},~
\bf  Minghui Liu \textsuperscript{b},~
\bf J\"{u}rgen Ulbricht \textsuperscript{c}~

\vskip 0.3in

{\small {\it

\textsuperscript{a}School Of Physics and Technology, Wuhan University, Wuhan, Hubei, China \\
\vspace{0.25cm}
\textsuperscript{b} Chinese University of Science and Technology, USTC, Hefei, Anhui 230 029, China \\
\vspace{0.25cm}
\textsuperscript{c} Swiss Institute of Technology ETH Zurich, CH-8093 Zurich, Switzerland \\

}
}

\vskip 0.3in

{\bf Abstract}

\end{center}


\baselineskip=18pt \noindent



\noindent We used data mainly collected by OPAL experiment to test the signal significance of a parameter $\varepsilon$, which is equal to zero in SM. For total cross sections, we obtained no derivation of enough significance. But for differential cross sections, some derivation over 3$\sigma$ are observed, indicating that this contact interaction might exist.

\vskip 3mm

\vskip 5mm


\section{Introduction}
$~$

Standard Model(SM) has achieved numerous number of success in explaining the phenomena of high energy physics and lead to very precise theoretical prediction in some topics. In SM, we have three generations of leptons, which are $e,~\mu,~\mathrm{and }~\tau$. In SM, all the leptons are assumed to be point-like particles. In the past few decades, many experiment tried to find the evidence of this assumption, and have also greatly investigated about the upper limit of electron size. 

However, many theorists believe that SM is only an effective theory which is only valid in a certain range. Discrepancy between experimental data and SM prediction were observed by the data that several detectors on LEP collected many years ago. Some models including a contact interaction may be able to explain the discrepancy. In this paper we conducted a data driven way to test a Bhabha scattering process
\begin{gather}
    \BHABHAG
\end{gather}
and tried to find the minimum interaction range of QED. Similar studies were conducted almost twenty years ago. Since now we have larger data, and better tools to calculate theoretical prediction, we tried to better extend this study. We used data collected by OPAL, L3, ALEPH, and DELPHI for this study.

\section{Contact Interactions}

Contact interactions is referred to as the interaction of particles within a very short distance. It will lead to varied results compared to normal interaction caused my basic interaction in an energy range $\Lambda>>\sqrt{s}$.

For fermion-pair production, the lowest order flavor-diagonal and helicity-conserving operators have dimension. The equation can be derived by extending Lagrangian. For simplicity, differential cross section equations can be discribed as follow\cite{PhysRevLett.50.811,opal1998tests, alexander1996test}:
\begin{align}
    &\frac{\dd \sigma}{\dd(\cos\theta)}=\frac{\pi\alpha^2}{4s}[4A_{0}+A_{-}(1-\cos\theta)^{2}+A_{+}(1+\cos\theta)^{2}];\label{diffCros}
    \\&A_{0}=\frac{1}{2}\left(\frac{s}{t}\right)^{2}\left|\mathbf{1}+\frac{g_{\mathrm{R}}g_{\mathrm{L}}}{e^{2}} \frac{t}{t_{z}}+\frac{\eta_{\mathrm{RL}}t}{\alpha\Lambda^{2}}\right|^{2}
     + \frac{1}{2}\left(\frac{s}{t}\right)^{2}\left|\mathbf{1}+\frac{g_{\mathrm{R}}g_{\mathrm{L}}}{e^{2}} \frac{t}{t_{z}}+\frac{\eta_{\mathrm{LR}}t}{\alpha\Lambda^{2}}\right|^{2},\quad 
    \\&A_{-}=\frac{1}{2}\left|\mathbf{1}+\frac{g_{\mathrm{R}}g_{\mathrm{L}}}{e^{2}} \frac{s}{s_{z}}+\frac{\eta_{\mathrm{RL}}s}{\alpha\Lambda^{2}}\right|^{2} +\frac{1}{2}\left|\mathbf{1}+\frac{g_{\mathrm{R}}g_{\mathrm{L}}}{e^{2}} \frac{s}{s_{z}}+\frac{\eta_{\mathrm{LR}}s}{\alpha\Lambda^{2}}\right|^{2},
    \\&A_{+}=\frac{1}{2}\left|\mathbf{1}+\frac{s}{t}+\frac{g_{\mathrm{R}}^{2}}{e^{2}}\left(\frac{s}{s_{z}}+\frac{s}{t_{z}}\right)+\frac{2\eta_{\mathrm{RR}}s}{\alpha\Lambda^{2}}\right|^{2}
+\frac{1}{2}\left|\mathbf{1}+\frac{s}{t}+\frac{g_{\mathrm{L}}^{2}}{e^{2}}\left(\frac{s}{s_{z}}+\frac{s}{t_{z}}\right)+\frac{2\eta_{\mathrm{LL}}s}{\alpha\Lambda^{2}}\right|^{2}
\end{align}
where $s$ is the square of energy in central mass frame. And the other variables are:

\begin{gather}
    t = -\frac{s(1-\cos\theta)}{2}\\
    s_z = s - \mu_Z^2 + i \mu_Z\Gamma_Z\\
    t_z = t - \mu_Z^2 + i \mu_Z\Gamma_Z \\
    \frac{g_L}{e} =  \tan\theta_W \\
    \frac{g_R}{e} =  -\cot2\theta_W
\end{gather}
and the differential cross section \cref{diffCros} can be express in the form:
\begin{equation}
    \frac{d\sigma}{d\Omega}=SM(s,t)+\varepsilon\cdot C_{Int}(s,t)+\varepsilon^2\cdot C_{CI}(s,t)
\end{equation}
where the first term is the Standard Model contribution, the second comes from interference between the SM and the contact interaction, and the third is the pure contact interaction effect. As a convention, $\frac{g^2}{4\pi} =1 $, and \cite{PhysRevD.64.071701}
\begin{gather}
    \varepsilon=\frac{g^2}{4\pi}\frac{\mathrm{sgn}(\eta)}{\Lambda^2}
\end{gather}

\section{Experimental Data}
All the data we use have been published. We mainly used the data collected by OPAL, which provided 105 bins of differential cross section in total 7 energy points from 189GeV to 207GeV, with the angle range of $-0.9<\cos\theta<0.9$ and accolinearity less than 10$^\circ$\cite{2004}. The data collected by DELPHI\cite{abreu2000measurement} and L3\cite{acciarri2000measurement} covered a angle range of $44^\circ<\cos\theta<136^\circ$, requiring acollinearity $\zeta<25^\circ$. And some other data. Here $\theta$ is the angle between the outgoing $e^+e^-$ and the incoming $e^-$ beam. Both $e^+~\mathrm{and}~e^-$ are required to be in this range if not point out specifically. 

\section{Analysis Method}
To give the theoretical prediction of SM in differential cross sections and total cross sections in various bins, we employed a Monte Carlo generator Babayaga@NLO\cite{calame2019status}. The phase space we use in the theoretical calculation is identical to the experimental data we use.

In this study, we took totally 125 differential bins and a certain number of total cross sections into account. And the way we test the derivation was to construct a $\chi^2$ defined as follows:
\begin{gather}
    \chi^2 = \sum(\frac{(\mathrm{Theo. Prediction}(SM,\varepsilon) -{\mathrm{Exp. Measurement} })^2}{\sigma_{\mathrm{Stat.}}^2 + \sigma_{\mathrm{Sys.}}^2})
\end{gather}
where  Theo. Prediction is the theory prediction of cross section after taking $\varepsilon$ into account, while the Exp. Measurement of cross section. $\sigma_{\mathrm{Stat.}}$ and $\sigma_{\mathrm{Sys.}}$ are the statistical and systematic uncertainty respectively. For differential cross sections and total cross sections we sum up all the results respectively.

\section{Results}
We took the following models into account, and results are also listed in the table\cite{PhysRevD.64.071701}.
    \begin{table}[h]
        \centering
        \begin{tabular}{cccc}
        \hline
        & Amplitudes &  $\varepsilon$&  $\Lambda$ \\
             
           Model  & [$\eta_{LL},\eta_{RR},\eta_{LR},\eta_{RL},$] & (TeV$^{-2}$) &  (TeV) \\ \hline
                      $LL$  &[$\pm 1, 0, 0, 0$] & $0.00446\pm0.00071$  &  14.97
             \\ 
                      $RR$  &[$0, \pm 1, 0, 0$] &$0.00456\pm0.00073$ &  14.80
             \\   
                      $LR$  &[$0, 0, \pm 1, 0$] &$0.00553\pm0.00072$& 13.44
             \\   
             
                      $RL$  &[$0, 0, 0, \pm 1$] & $0.00553\pm0.00072$&  13.44
             \\   
             
                      $VV$  &[$\pm 1, \pm 1, \pm 1, \pm 1$] &$0.00133\pm0.00017$ &  27.42
             \\   
             
                      $AA$  &[$\pm 1, \pm 1, \mp 1, \mp 1$] & $0.00139\pm0.00045$&  27.14
             \\   
                      $LL-RR$  &[$\pm 1, \mp 1, 0, 0$] & $0.00034\pm0.00103$&  31.16
             \\   
             
                      $LL+RR$  &[$\pm 1, \pm 1, 0, 0$] & $0.00225\pm0.00103$&  21.08
             \\   
             
                      $LR+RL $ &[$0, 0 , \pm 1, \mp 1 $] & $0.00309\pm0.00040$&  17.98
             \\   \hline

        \end{tabular}
        \caption{Models And Results}
        \label{tab:my_label1}
    \end{table}
    
For now the results only includes statistical uncertainty.

\FloatBarrier

\appendix
\section*{Appendix}
\section{Fitting results}

\begin{figure}[htbp]
    \centering
    \includegraphics[width=0.8\linewidth]{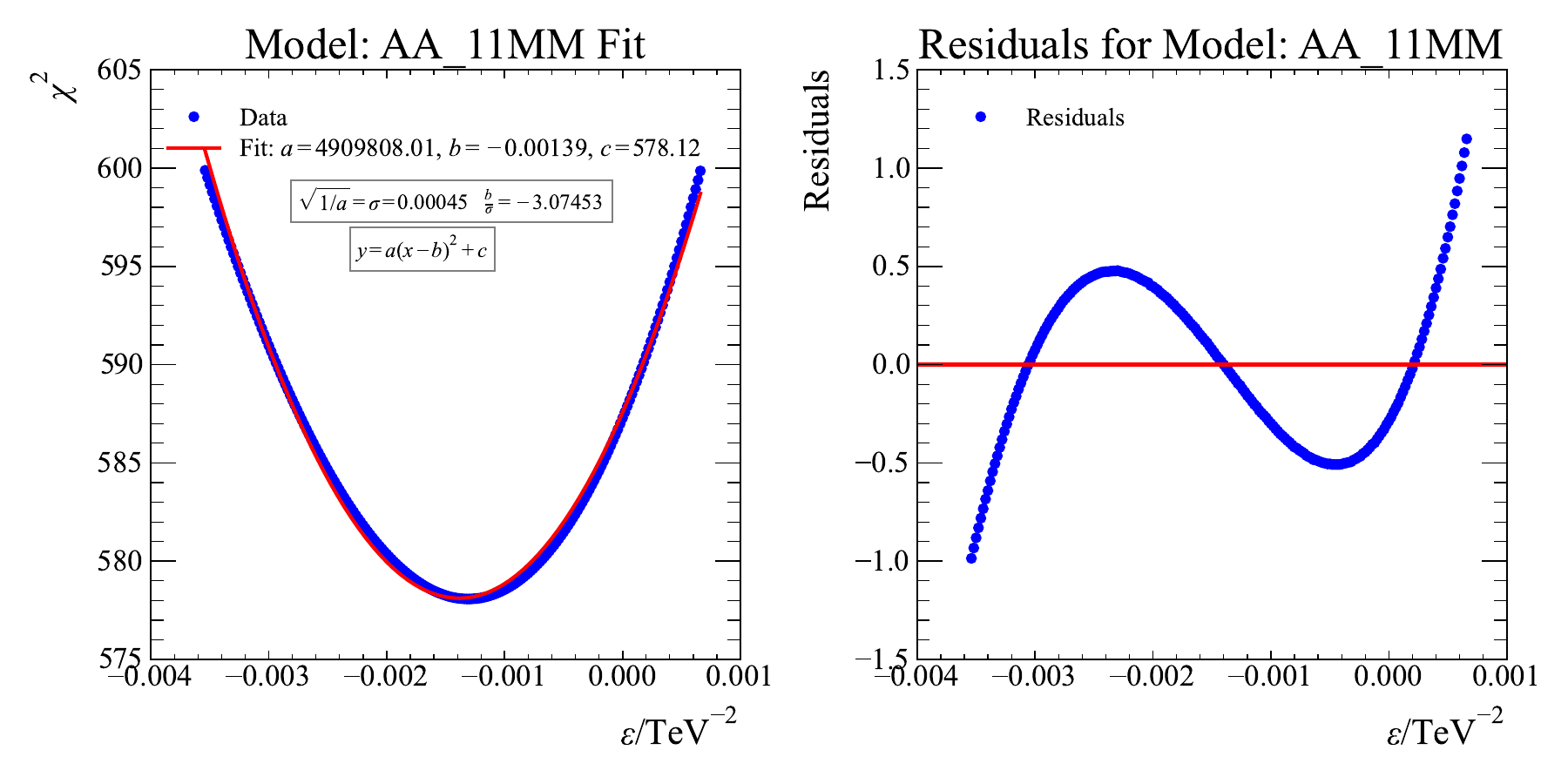}
    \caption{Model AA}
    \label{fig:enter-label2}
\end{figure}

\begin{figure}[htbp]
    \centering
    \includegraphics[width=0.8\linewidth]{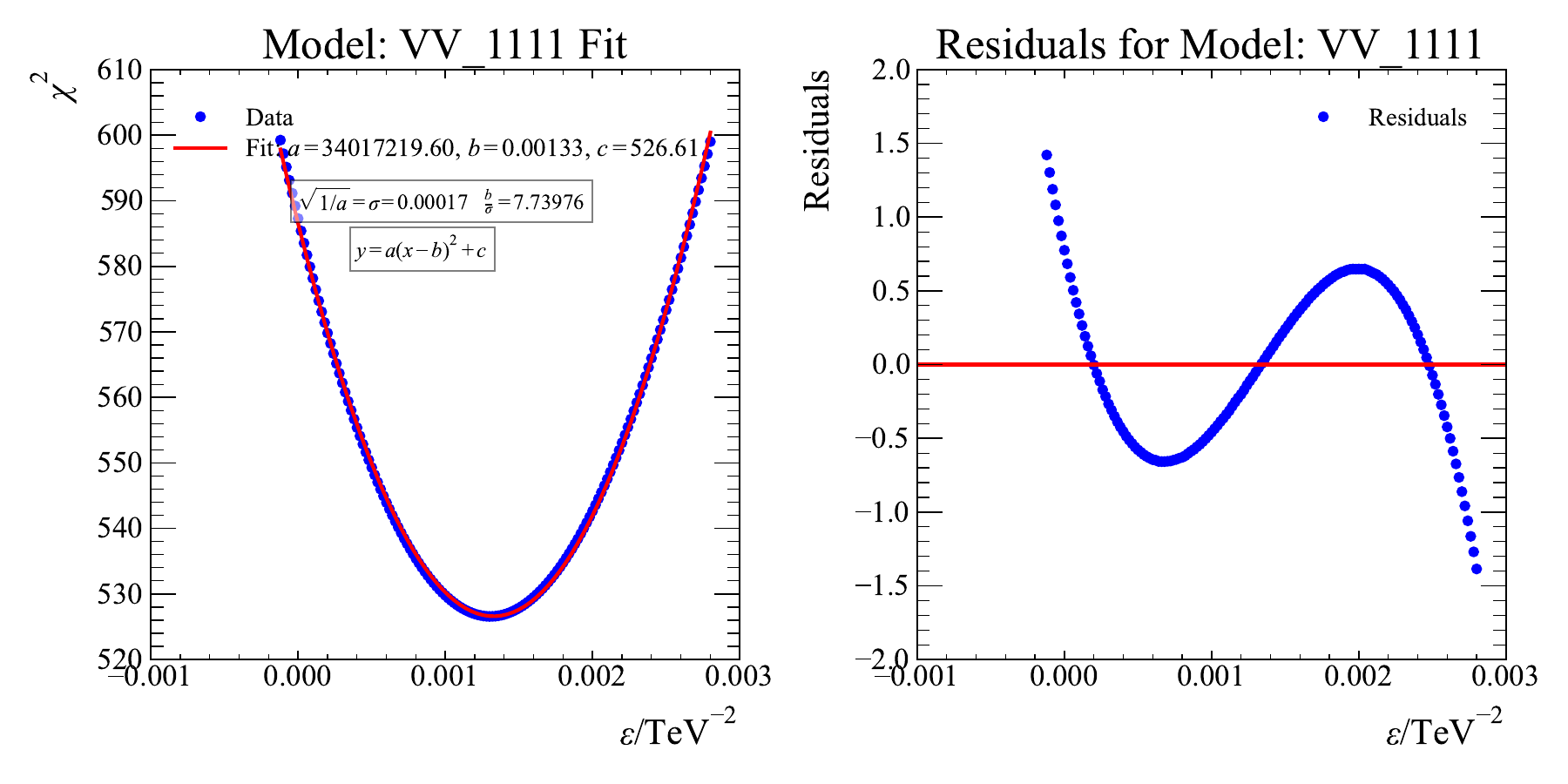}
    \caption{Model VV}
    \label{fig:enter-label3}
\end{figure}

\begin{figure}[htbp]
    \centering
    \includegraphics[width=0.8\linewidth]{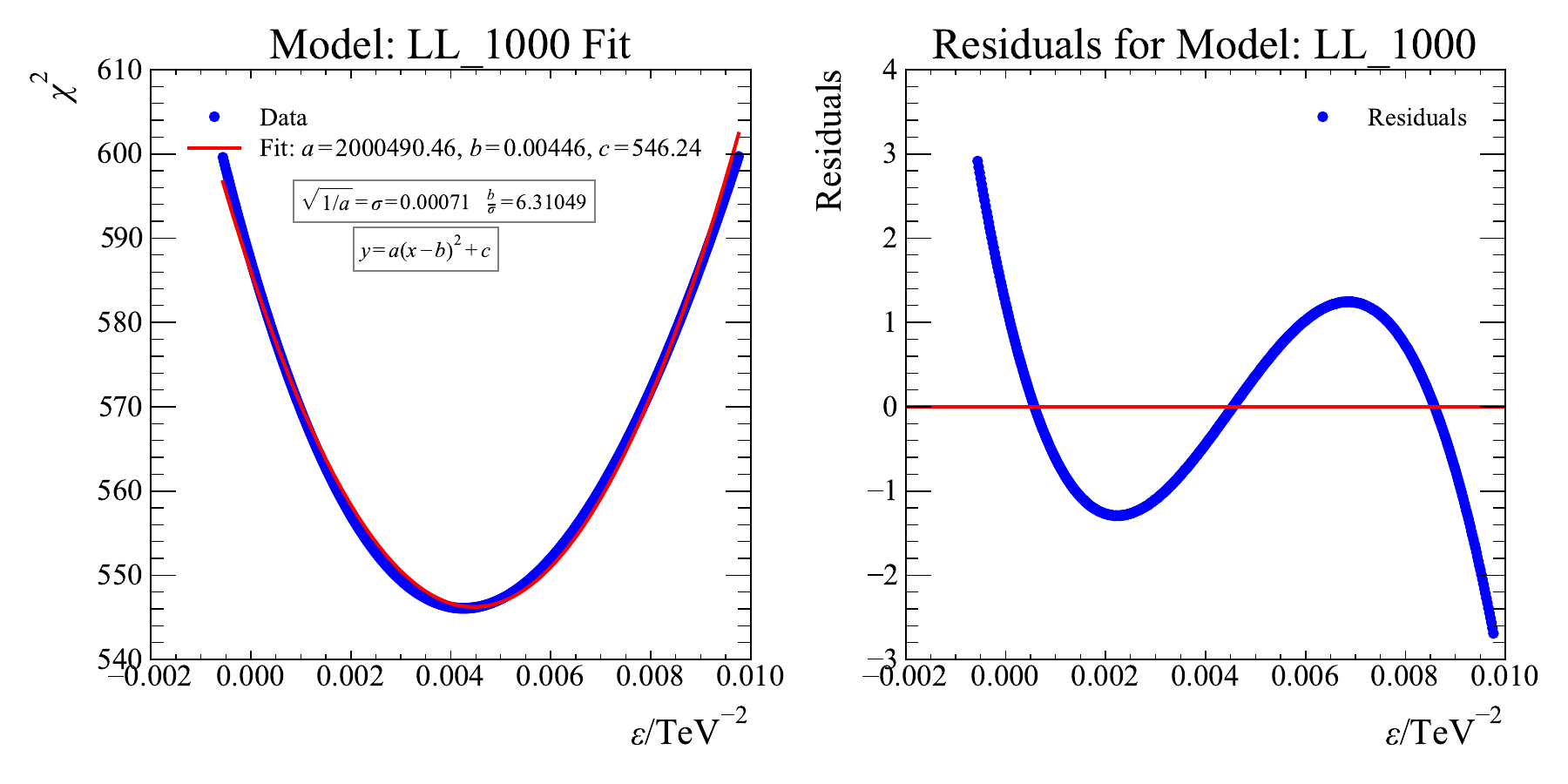}
    \caption{Model LL}
    \label{fig:enter-label4}
\end{figure}

\begin{figure}[htbp]
    \centering
    \includegraphics[width=0.8\linewidth]{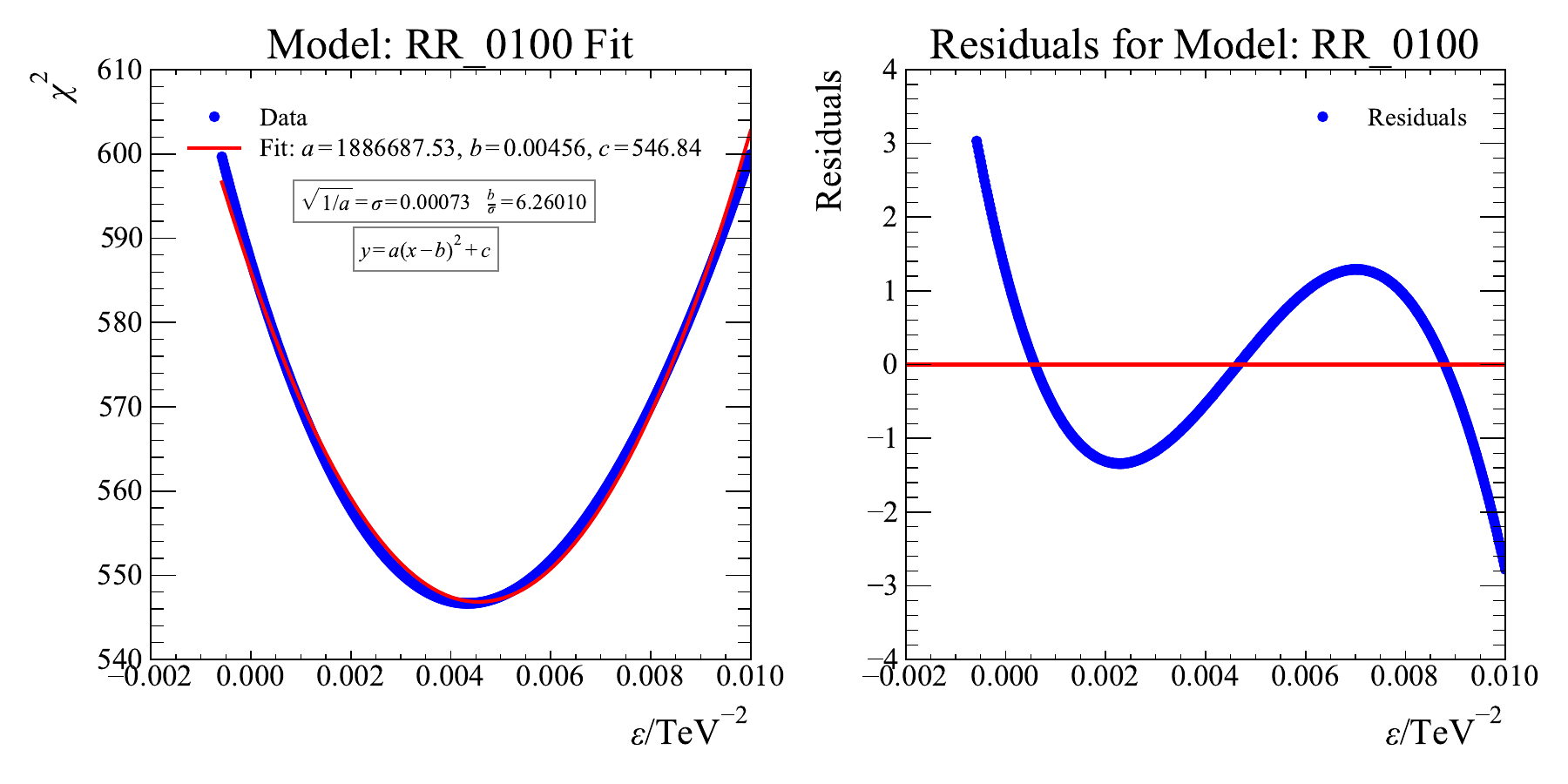}
    \caption{Model RR}
    \label{fig:enter-label5}
\end{figure}

\begin{figure}[htbp]
    \centering
    \includegraphics[width=0.8\linewidth]{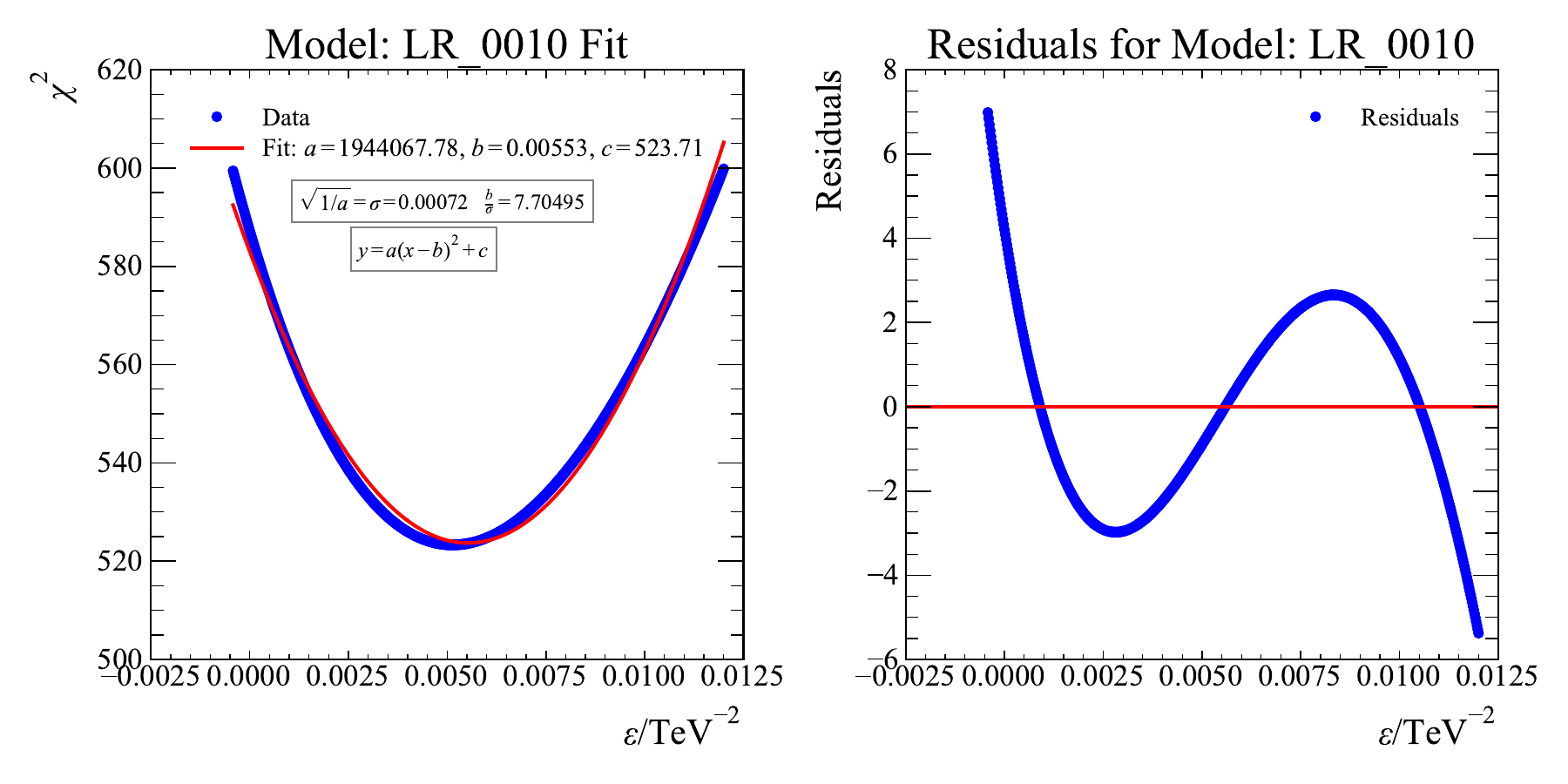}
    \caption{Model LR}
    \label{fig:enter-label6}
\end{figure}

\begin{figure}[htbp]
    \centering
    \includegraphics[width=0.8\linewidth]{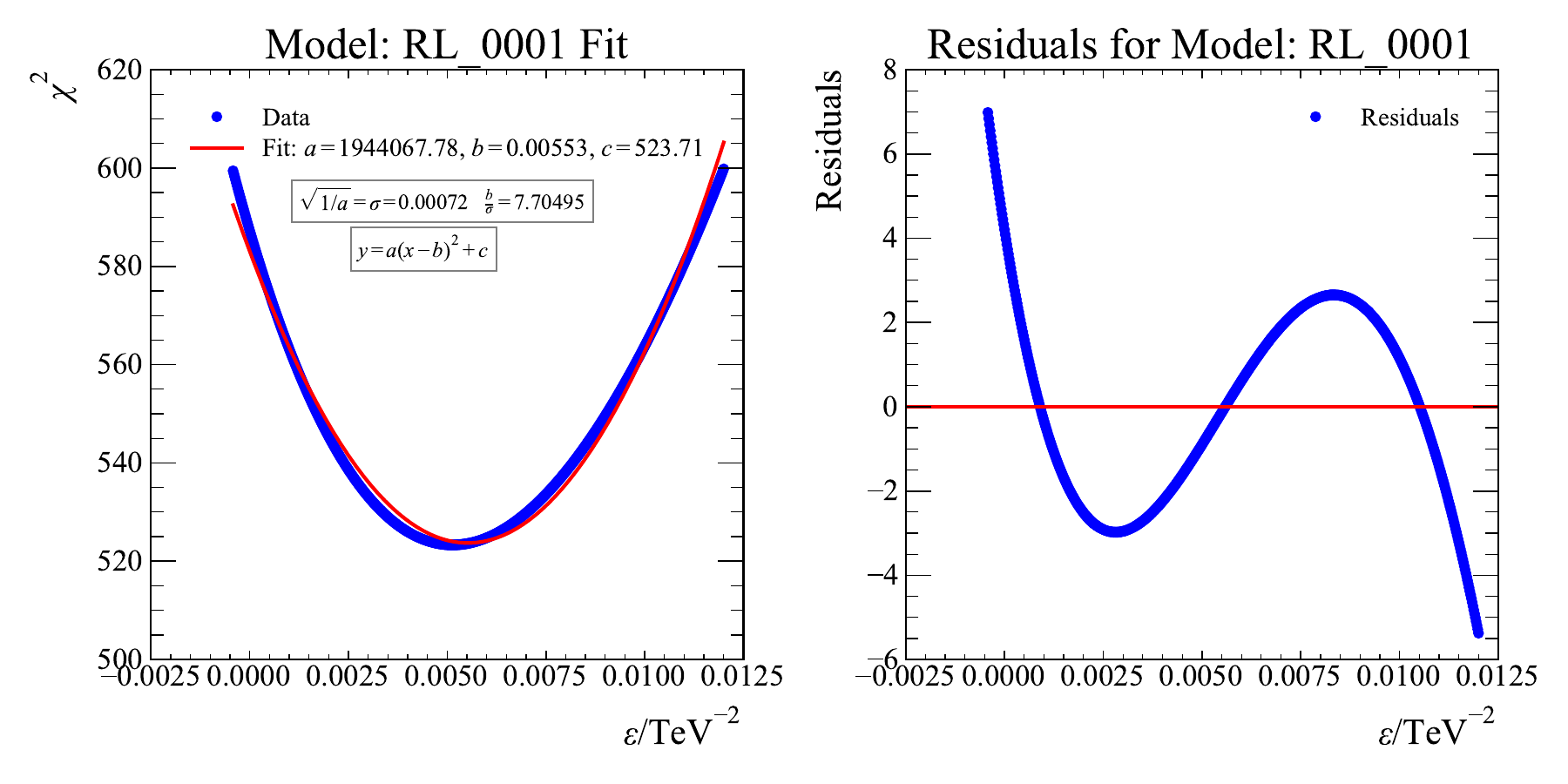}
    \caption{Model RL}
    \label{fig:enter-label7}
\end{figure}

\begin{figure}[htbp]
    \centering
    \includegraphics[width=0.8\linewidth]{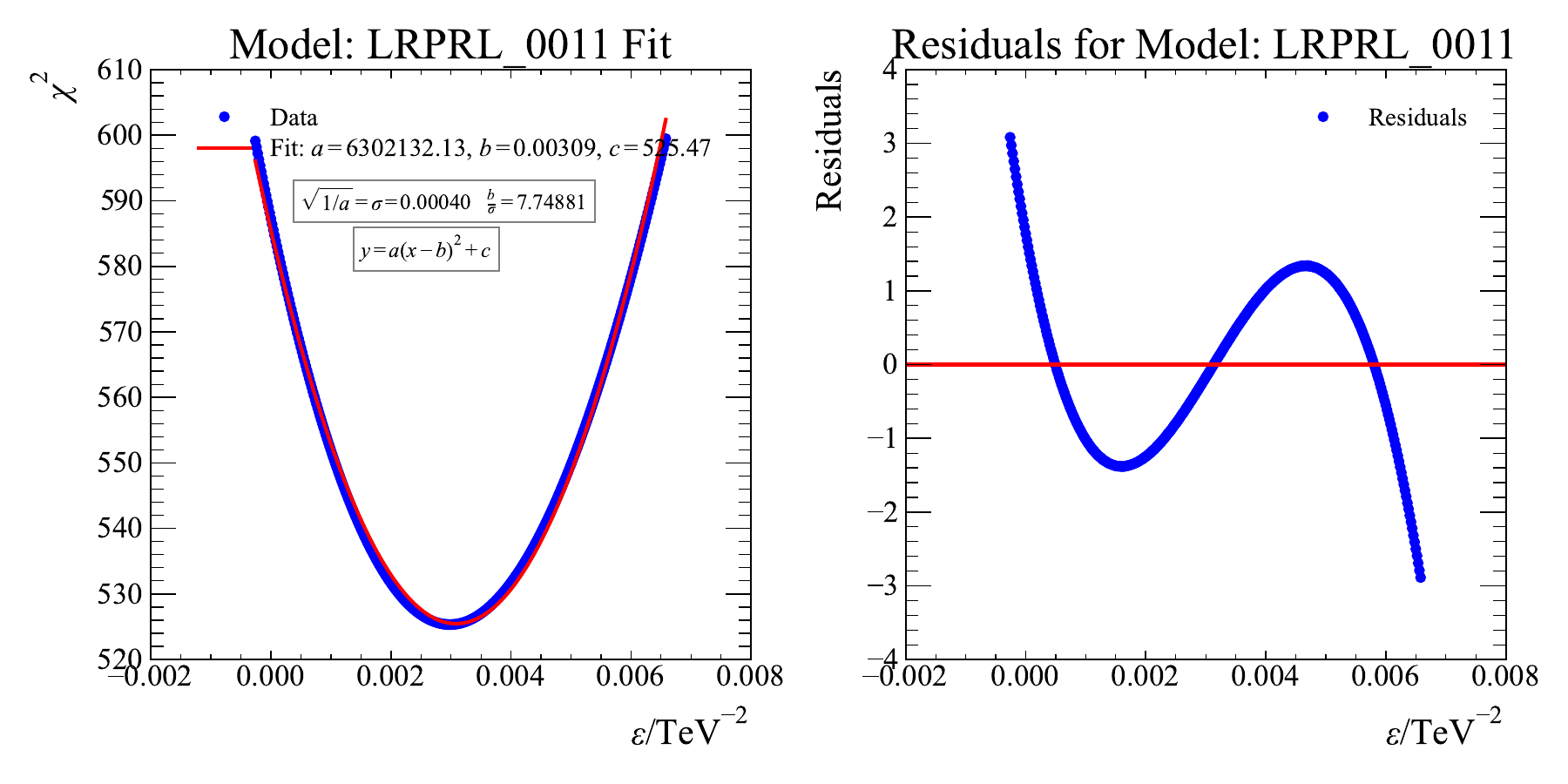}
    \caption{Model LR+RL}
    \label{fig:enter-label8}
\end{figure}

\begin{figure}[htbp]
    \centering
    \includegraphics[width=0.8\linewidth]{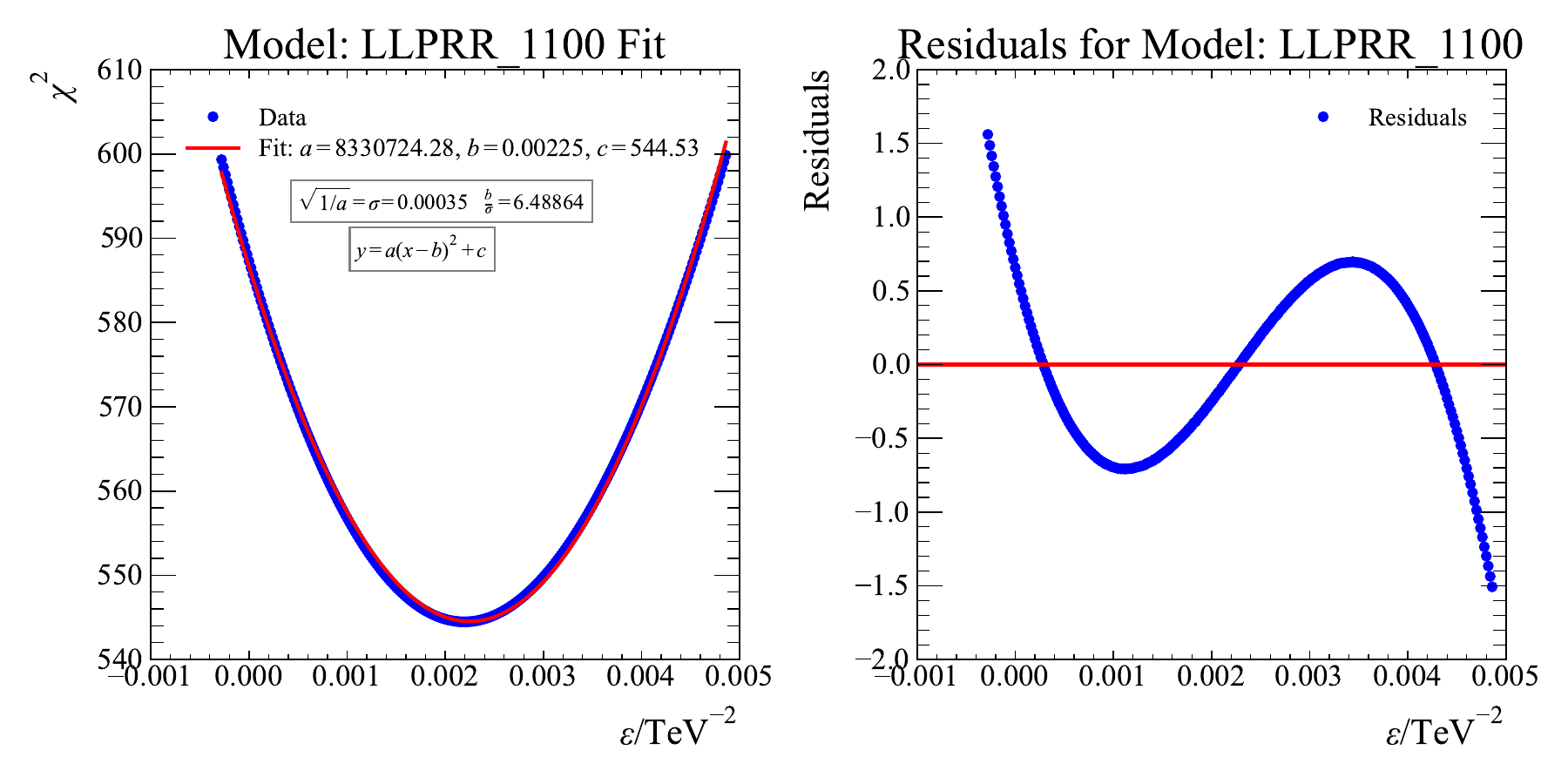}
    \caption{Model LL+RR}
    \label{fig:enter-label9}
\end{figure}

\begin{figure}[htbp]
    \centering
    \includegraphics[width=0.8\linewidth]{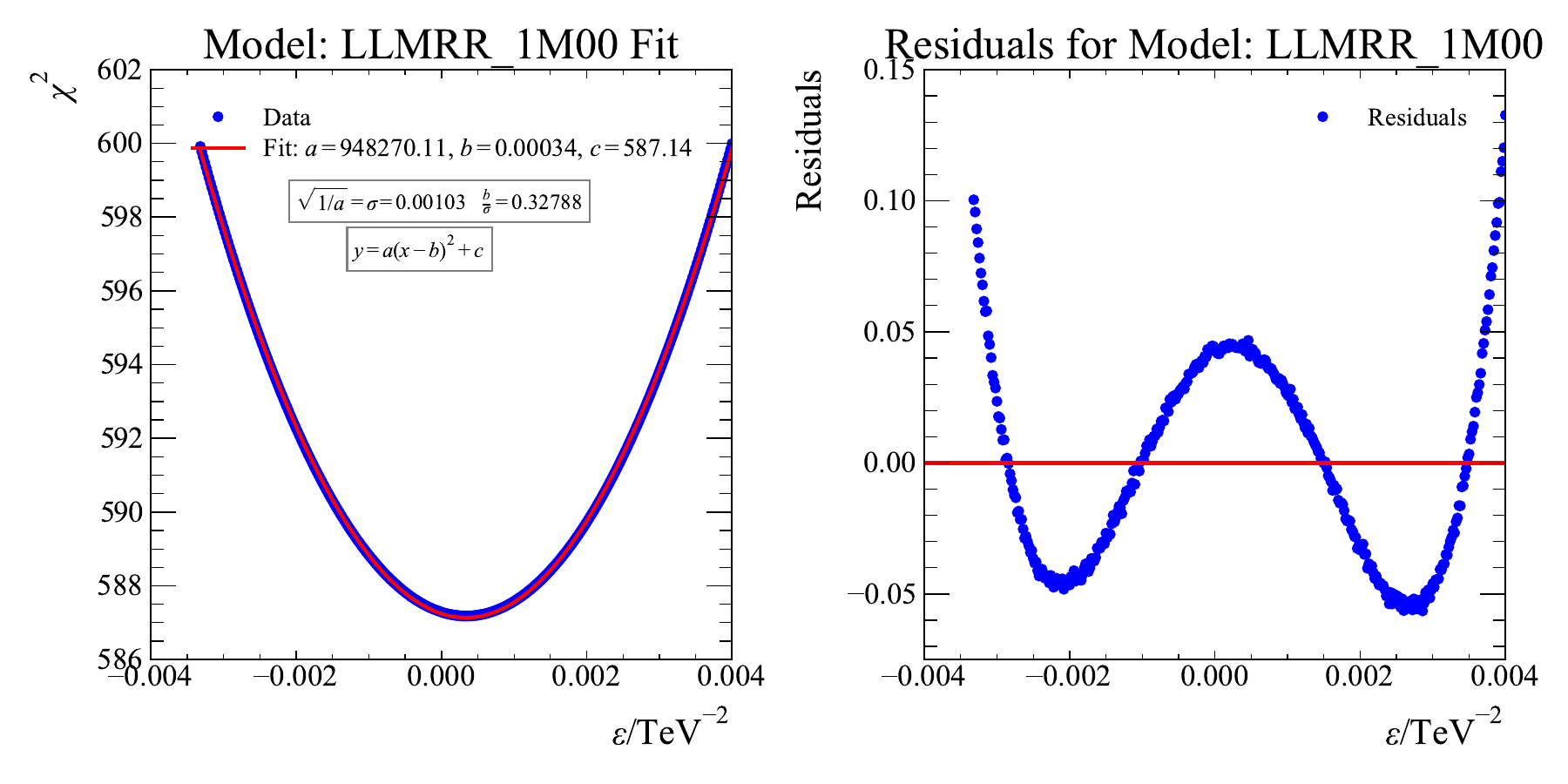}
    \caption{Model LL-RR}
    \label{fig:enter-label10}
\end{figure}

\clearpage
\bibliographystyle{plain}  
\bibliography{references}   
\end{document}